\documentstyle[11pt,epsf]{article}

\begin{document}
\baselineskip 18pt
\title{Stability of a hard-sphere binary quasicrystal}

\author{H. M. Cataldo}
\date{}
\maketitle

\begin{center}
{\it Departamento de F\'{\i}sica,
 Facultad de Ciencias Exactas y Naturales, \\
Universidad de Buenos Aires,
 RA-1428 Buenos Aires, Argentina\\
and Consejo Nacional de Investigaciones
Cient\'{\i}ficas y
T\'ecnicas, Argentina}
\end{center}

\hspace{1cm}
\begin{center}
ABSTRACT
\end{center}
The stability of a  quasicrystalline structure, recently obtained
in a molecular-dynamics simulation of rapid cooling of a binary melt,
is analyzed for binary hard-sphere mixtures
within a density-functional approach.
It is found that this quasicrystal is metastable relative to crystalline
and fluid phases for diameter ratios above 0.83. Such trend is partially
reversed for lower diameter ratios, since the quasicrystal becomes stable
with respect to the crystal but does not reach a coexistence with the fluid.

\newpage

\section*{\S1. INTRODUCTION}

It is well known that quasicrystalline phases
 are obtained from alloys
of at least two metallic elements, mostly by processes of rapid
solidification of the melt. However, the question of how quasicrystals
form is far to be understood from the theoretical viewpoint (Steinhardt
and Di Vincenzo 1991, Janot 1994).
 In this respect,
an important step towards this goal could be given by a systematical
study of
numerical simulations at which a quasicrystalline phase is formed from
a  binary fluid. Unfortunately, this program is difficult to carry out
since simulations of such systems are very expensive due to the large
number of atoms which are
necessary to display the quasiperiodic features. Recently, however,
it was reported a first effort of a molecular dynamics simulation of
quasicrystal solidification from the melt of a diatomic three-dimensional
system (Roth, Schilling and Trebin 1995).
In fact,
it was shown that a binary Lennard-Jones fluid, at adequate cooling rates,
indeed solidifies into a phase of icosahedral long-range order close to
a quasicrystalline structure known as the {\em truncated icosahedral binary
model} (TIBM). This is assumed to be a simplified version of the
Henley-Elser model
 for the icosahedral structure of (Al,Zn)$_{49}$Mg$_{32}$
(Henley and Elser 1986).
A previous investigation by means of relaxation simulations
had shown that the TIBM was at least strongly metastable for T=0, a feature
which was mainly interpreted due to the fact that such structure comprises a
very dense packing of spheres of two sizes (Roth, Schilling and Trebin 1990).

In the present paper we shall investigate the stability of
a TIBM composed by hard-spheres by means of a density-functional
theory (Lutsko and Baus 1990),
which was previously applied to a one-component
hard-sphere (HS) quasicrystal (Cataldo and Tejero 1995).
HS solids may serve
as suitable reference systems for mean-field- or perturbation-theory treatments of more
realistic potentials (Daanoun, Tejero and Baus 1994, Achrayah and Baus 1998).
Recently, Denton and Hafner (1997a,b)
have studied a model of monatomic
quasicrystal interacting via effective metallic pair potentials, by means of
a combination of classical density-functional theory and
 thermodynamic perturbation theory, which takes as its reference system the
above mentioned one-component HS quasicrystal.

Closely related to freezing and thermodynamic stability properties of HS solids
is the problem of packing. This is a well studied subject for identical hard
spheres, but less is known about packing properties of mixtures of spheres or
disks of two different sizes, see e. g., Likos and Henley (1993).

The possibility of quasicrystalline
metastable states for one- and two-component HS solids has been
recently
explored within density-functional theory in the range of diameter ratios
above 0.85 (M$^c$ Carley and Ashcroft 1994).
Then it was found that such phases have  higher
free energies than the crystalline or liquid states.
We shall see that a similar conclusion stands for the TIBM for diameter
ratios above 0.83.
However, this trend is partially reversed for lower diameter
ratios, since the quasicrystal becomes
stable with respect to crystal
but remains metastable with respect to fluid.

In the following Section we shall
describe the TIBM sphere packing, showing that a small shift in the original
position of large spheres makes the maximum packing fraction to become
competitive with respect to HS crystal structures of similar composition.
In Section 3 we shall give a brief account of the density-functional approach
to the calculation of quasicrystal free energy. Finally, our results and
conclusions will be presented in Sections 4 and 5, respectively.

\section*{\S2. TIBM SPHERE PACKINGS}
Let us  consider
a decoration of the 3D Penrose tiling at which all vertices and middle edges
of both rombohedral cells are decorated by identical spheres.
Thus, if we take rombohedral edges as unity, the maximum
diameter turns out to be 0.5. On the
other hand, each prolate rombohedron can accomodate two larger spheres
located symmetrically at both sides of the center of its long diagonal
(see fig. 1). If the location of such spheres divides the diagonal into
three parts of ratios $\tau^{-2}:\tau^{-3}:\tau^{-2}$, with $\tau=(\sqrt{5}+1)/2$, the
maximum diameter (given by the separation of the centers of both spheres)
is 0.563, yielding a diameter ratio 0.888 and a maximum packing
fraction of 0.580. Here it is convenient to define a notation for the
HS parameters namely, $\alpha=\sigma_1/\sigma_2$, $x_2=N_2/N$, and
$\eta=\frac{\pi}{6}[(1-x_2)\alpha^3+x_2]\frac{N}{V}\sigma_2^3$, denote
respectively, the diameter ratio, the concentration of large spheres, and
the packing fraction.

Now, taking into account that in the Henley-Elser model the above ideal
position of the larger spheres was chosen essentially for
simplicity (Henley and Elser 1986),
we may allow a greater separation between both
spheres in order to increase the maximum diameter. In fact, by this
procedure the maximum packing fraction can be improved to $\eta_{max}=
0.648$, for a diameter ratio $\alpha=0.798$ when large spheres of
adjacent rombohedra touch each other (in addition to the contact between
both large spheres of each rombohedron).

In order to measure the relevance of the above packing fractions, we may
consider a binary crystal of similar composition namely, a fcc lattice
plus a decoration of small spheres on faces and large spheres on vertices,
which has a concentration $x_2=0.25$ close to the TIBM value (0.236).
In addition, we may compare to a {\em disordered-fcc} structure in which
both kind of spheres are distributed irregularly over the sites of a fcc
lattice. This disordered crystal allows us to select the exact composition
of the TIBM.

In table \ref{tab1} we give the packing fractions of the above structures
for three selected values of the diameter ratio. We note that the diameter
ratio 0.798 yields the better quasicrystal performance with a packing fraction
slightly higher than the crystal one (the packing fraction of the disordered
crystal is only relevant at the higher diameter ratio).

\renewcommand\arraystretch{2}
\tabcolsep 1.5cm
\begin{table}
\caption{\baselineskip 18pt
 Maximum packing fraction of the TIBM quasicrystal (q),
the crystal (c), and the disordered crystal (d), for three selected values
of the diameter ratio  $\alpha$.
\label{tab1}}
\begin{tabular}{cccc} \hline\hline

$\alpha$ & q & c & d\\
\hline
0.888 & 0.580 & 0.683 & 0.572\\
0.833 & 0.618 & 0.658 & 0.502\\
0.798 & 0.648 & 0.643 & 0.462\\ \hline\hline
\end{tabular}
\end{table}

\section*{\S3. CALCULATION OF THE HELMHOLTZ FREE ENERGY}
The above calculation of packing fractions may be regarded as an
``order-zero" of a stability analysis of the competing HS
structures. In order to get a much more definite analysis, in what follows
we shall resort to a density-functional approach, the so-called
{\em generalized effective liquid approximation} (Lutsko and Baus 1990),
which has proven to
provide a very accurate description of one-component HS solids.
In particular, the application of this formalism to a monatomic HS
quasicrystal of high packing fraction (Henley 1986), was recently
reported (Cataldo and Tejero 1995),
showing that it is thermodynamically unstable
(i. e., it does not have a coexistence with the fluid),
in agreement with previous studies (M$^c$ Carley and Ashcroft 1994).
Thus, our present treatment will be based upon a straightforward
generalization to mixtures of the above approach. In fact, following
Xu and Baus (1992), we write the Helmholtz free energy per particle of
a HS  binary solid as the following functional of both partial local number
densities $\rho_i({\bf r})$ ($i=1,2$),
\begin{equation}
\label{1}
f[\rho_1,\rho_2]= \frac{k_BT}{N}\sum_{i=1}^2\int d {\bf r}
\rho_i({\bf r })[\ln(\Lambda_i^3\rho_i({\bf r }))-1]+f_{ex}[\rho_1,\rho_2],
\end{equation}
where $k_B$ denotes Boltzmann's constant and $\Lambda_i$ the thermal de
Broglie wavelenght of species $i$. The excess free energy per particle of
the solid $f_{ex}$ is approximated by the one of an {\em effective} HS
binary fluid of the same concentration, $f_{ex}/k_BT\simeq\psi
(\hat{\eta})$, whose packing fraction $\hat{\eta}$ is obtained
from the solution of a system of two coupled differential
equations (Tejero and Cuesta 1993),
\label{20}
\begin{equation}
\label{21}
\hat{\eta}^{\prime}(\lambda)
= \frac{z(\lambda)-\psi(\hat{\eta}(\lambda))}
{\lambda\psi^{\prime}(\hat{\eta}(\lambda))}
\end{equation}

\begin{equation}
\label{22}
 z^{\prime}(\lambda)=\Phi(\hat{\eta}(\lambda))
\end{equation}
as $\hat{\eta}=\hat{\eta}(\lambda=1)$. In the above equations, the prime
denotes function derivative, the initial conditions are $\hat{\eta}(0)=
z(0)=0$, and
\begin{equation}
\label{23}
\Phi(\hat{\eta}(\lambda))= -\frac{1}{N}\int d{\bf r} \int d{\bf r'}
\sum_{i=1}^{2}\sum_{j=1}^{2}\,\rho_i({\bf r})\,
\rho_j({\bf r'})\,C_{ij}^{0}(|{\bf r}-{\bf r'}|,\hat{\eta}(\lambda)),
\end{equation}
where $C_{ij}^{0}$ denotes the Percus-Yevick approximate direct correlation
function of a binary HS fluid mixture (Lebowitz 1964).
The HS density profiles
are parametrized in terms of Gaussians centered at the sites
$\{{\bf R}^{(i)}\}$ occupied by species $i$:
\begin{equation}
\label{24}
\rho_i({\bf r}) = \left(\frac{\alpha_i}{\pi}\right)^{3/2}\sum_{{\bf
R}^{(i)}}
\mbox{e}^{-\alpha_i({\bf r}
-{\bf R}^{(i)})^2}\;\;\;\;\;\;\;i=1,2.
\end{equation}
Replacing (\ref{24}) in (\ref{23}), and performing the angular integrations,
we obtain\footnote
{In case of a disordered structure, instead of (\ref{24}) we must
consider the alternative expression,
$\rho_i({\bf r}) = x_i\left(\frac{\alpha_i}{\pi}\right)^{3/2}\sum_{{\bf
R}}
\exp[-\alpha_i({\bf r}
-{\bf R})^2]\;\;(i=1,2)$,
where the summation runs over {\em all} lattice sites. Accordingly, the
expression (\ref{27}) has to be modified by the replacements,
$\sum_{{\bf R}^{(i)}}^{N_i}\rightarrow x_i\sum_{{\bf R}}^N$.}
\begin{equation}
\label{27}
\Phi(\hat{\eta}(\lambda))= -\frac{1}{N}
\sum_{i=1}^{2}\sum_{j=1}^{2} \sum_{{\bf R}^{(i)}}^{N_i}
\sum_{{\bf R}^{(j)}}^{N_j}
\int_0^{\infty}\, dr\,r\, C_{ij}^{0}
(r,\hat{\eta}(\lambda)) S(r,\alpha_{ij},r_{ij}),
\end{equation}
being
\begin{equation}
\label{28}
S(r,\alpha_{ij},r_{ij}) =\left[\frac{\alpha_{ij}}{2\pi r_{ij}^2}\right]^{1/2}
\left[ \exp(-\alpha_{ij}(r-r_{ij})^2/2)-\exp(-\alpha_{ij}(r+r_{ij})^2/2)
\right],
\end{equation}
where $\alpha_{ij}=2\alpha_i\alpha_j/(\alpha_i+\alpha_j)$ and
$r_{ij}= |{\bf R}^{(i)}-{\bf R}^{(j)}|$.
The calculation of (\ref{27}) deserves some discussion. First, we note that the
integral in $r$ can be analytically computed and, in the case of {\em periodic}
structures, the double lattice sum reduces to a simple one running over
spherical shells of sites centered around an arbitrary site. On the other
hand, the lack of periodicity of the quasicrystal does not allow this
simplification, making necessary to resort to another method. In a
previous work (Cataldo and Tejero 1995),
it was proposed a  method of evaluation of quasilattice
sums that can be applied to the present case. We refer the reader for
details to such work, but in a brief summary  the method
consists in a generalization of the above mentioned shell summations of
periodic lattices to a situation in which each site produces a different
``shell" sum due to variations in the surroundings, as is the case of a
quasilattice. So, the double sum consisting in the average of such ``shell"
sums over an {\em infinite} number of sites is approximated by an average
over the sites within a finite size quasicrystal ``sample''.
Here it is important to notice that in the case of a crystal, this
procedure yields the correct value only
if the ``sample'' possesses the exact
composition of the infinite crystal (note that it is subjected to
variations due to boundaries). So, we have adopted this {\em exact
composition} prescription for the quasicrystal ``samples''.

The Gaussian width parameters $\{\alpha_1,\alpha_2\}$ of (\ref{24}) are
determined so as to minimize the total free energy of (\ref{1}).
 Fig. 2  shows the effect of quasilattice sums on the value
of quasicrystal free energy; we see that it suffices to work with ``samples''
of $N_2>1000$ large spheres in order to reduce convergence errors
to less than 0.1$\%$. In fact, the computation time needed for evaluating
each ($\alpha_1,\alpha_2$) free energy value is around 11' CPU in a
workstation Alpha DEC 3000 for a ``sample'' of 4000 large spheres.
Fortunately, the {\em location} of free energy minima in the
$\alpha_1,\alpha_2$ plane is rather insensitive to the ``sample'' size,
so we could work with small ``samples'' ($N_2\sim 100$) in locating
minima and, at a second stage, a minimum value of free energy was
computed for a large ``sample'' ($N_2\sim 4000$) within a small grid of
typically 20 ($\alpha_1,\alpha_2$) points.

An alternative treatment of the problem of quasilattice sums consist in
replacing the quasicrystal by a suitable crystalline approximant
(Kraj\v{c}\'{\i} and Hafner 1992, Goldman and Kelton 1993). Such structures
are crystals with large unit cells where the arrangements of sites closely
approximate the local structure of the quasicrystal. Instead of an aperiodic
tiling of space by prolate and oblate rombohedra (Penrose tiling),
approximants are based upon tilings by the same rombohedral units which are
periodically repeated in the form of large cubic unit cells. Once such
tiling is defined, the decoration of both kind of rombohedra with large and
small spheres can be done in the same way as in the quasicrystal. Lattice sums
over the approximant are then computed by taking any of its cubic unit cells
as a ``sample" in the above explained sense. It is clear that approximants
having larger unit cells are expected to yield free energy values closer to
the quasicrystal ones. However, for HS potentials only the most local
structure in the neighbouring of each particle should be of importance, and
then one expects that the smallest approximants would be adequate to obtain
acceptable results. We have found that this is indeed the case; one of the
smallest approximants is the so-called `1/1' cubic structure (Elser and
Henley 1985), which has 20 prolate and 12 oblate rombohedra per unit cell.
It is slightly richer in large spheres (+0.84\%) than the quasicrystal and
thus has a 0.03\% (0.11\%) higher packing fraction for $\alpha$=0.888
($\alpha$=0.798). Regarding its free energy, it turns out to be always
slightly lower than the quasicrystal one, namely, for $\alpha$=0.888 the
free energy differences vary from 0.13\% (nearly within convergence errors
of the quasicrystal values)
 at low packing fractions, up to 0.21\% at high packing fractions, while
for $\alpha$=0.798, the values increase from 0.21\% up to 0.41\%.
From all these figures, it becomes clear that such differences in free
energy are most probably due to the slightly higher packing fraction of
the 1/1 approximant.

\section*{\S4. RESULTS}
In fig. 3 we represent the free energy of the different phases for
$\alpha=0.888$, note that the quasicrystal turns out to be always
metastable with respect to the remaining phases. On the
other hand, both crystals show an exchange of relative stability with
respect to
the fluid\footnote
{The fluid phase free energy was computed from the BMCSL
equation (Boublik 1970,
Mansoori,  Carnahan, Starling and Leland 1971).},
which is consistent with a freezing into the
disordered solid as expected (M$^c$ Carley and Ashcroft 1994,
Xu and Baus 1992).

Another feature of solid phases
concerns to the degree of localization of the spheres at the corresponding
sites. This is measured by the Gaussian widths which are shown in table
\ref{tab2}. Note the particular behavior of small spheres in the
quasicrystal, they present an appreciably lower localization than the
large or small spheres of the remaining cases.

\tabcolsep 0.6cm
\begin{table}
\caption{\baselineskip 18pt
Gaussian width $(\alpha_i)^{-\frac{1}{2}}$ in units of the
respective HS diameter $\sigma_i$ (diameter ratio
 $\alpha=0.888$), for the quasicrystal (q), the crystal (c), and the
disordered crystal (d).
\label{tab2}}
\begin{tabular}{ccccccc}    \hline\hline
$\eta$ & $(\alpha_1)_q^{-\frac{1}{2}}$
& $(\alpha_2)_q^{-\frac{1}{2}}$  & $(\alpha_1)_c^{-\frac{1}{2}}$  &
$(\alpha_2)_c^{-\frac{1}{2}}$  &
$(\alpha_1)_d^{-\frac{1}{2}}$  & $(\alpha_2)_d^{-\frac{1}{2}}$ \\
\hline
0.52 & 0.20 & 0.11 & 0.12 & 0.11 & 0.13 & 0.14 \\
0.53 & 0.17 & 0.10 & 0.11 & 0.094 & 0.12 & 0.12 \\
0.54 & 0.16 & 0.093 & 0.10 & 0.085 & 0.11 & 0.11 \\
0.55 & 0.15 & 0.086 & 0.097 & 0.078 & 0.10 & 0.10 \\
0.56 & 0.14 & 0.080 & 0.090 & 0.071 & 0.092 & 0.094 \\
0.57 & 0.13 & 0.075 & 0.082 & 0.064 & 0.085 & 0.090 \\
0.58 & 0.12 & 0.070 & 0.076 & 0.058 &  &  \\
0.59 &     &      & 0.070 & 0.053 &  &  \\
0.60 &     &      & 0.064 & 0.048 &  &  \\
0.61 &     &      & 0.059 & 0.044 &  & \\
0.62 &     &      & 0.054 & 0.039 &  &  \\
0.63 &     &      & 0.049 & 0.036 &  &  \\  \hline\hline
\end{tabular}
\end{table}

An additional test on quasicrystal stability may be performed by taking
the distance $D$ between the centers of
both large spheres of each rombohedron as a
variational parameter. In fact, in fig. 4 we find free energy
minima for $D$ values which are approximately equidistant from the
original $D$ in the Henley-Elser model
and the one that allows to
maximize the packing fraction (both distances are indicated by dotted
vertical lines). It is interesting to note that the increase of free energy
at both sides of minima is accompanied by a loss of particle
(small and large spheres) localization.
Moreover, for each packing fraction there exists a critical maximum
separation (star points in fig. 4)  beyond which the free energy
minimum in $\alpha_1,\alpha_2$ vanishes, yielding thus  an end of mechanical
stability. In order to understand such
 behavior, we first note from  fig. 4 that the critical $D$
increases with $\eta$, and then for a fixed $D$, say 0.66, the quasicrystal
becomes mechanically unstable below certain density, $\eta\simeq 0.53$ in
this case. In other words, the bounds of mechanical stability in
fig. 4 denoted as star points, were to be expected as arising
from the familiar fact that any solid becomes unstable below certain
density.

Let us now repeat our analysis for the lower diameter ratio $\alpha=0.798$.
Fig. 5 represents the free energy of quasicrystal (for $D=0.627$),
crystal, and fluid. We see that the improved quasicrystal packing
fraction has contributed to
stabilize such phase with respect to the crystal, but it
remains unstable as compared to the fluid. Note that the crystal free
energy is very close to the quasicrystal one at $\eta=0.53$, but
unfortunately it was impossible to perform a reliable calculation of the
crystal free energy for $\eta<0.53$ since the $\alpha_1,\alpha_2$
minimum at those low densities began to suffer an overlap with a spurious
minimum arising from the flaw of Percus-Yevick approximate direct
correlation function of the effective fluid at solid-like
densities\footnote
{Such spurious minima were discussed  by Lutsko and Baus (1990) for
one-component HS crystals.}.
On the other hand, we have found that the crystal
 minimum in $\alpha_1,\alpha_2$  vanishes for $\eta>0.59$, this unexpected
behavior seems to be caused by some kind of high-density flaw
of our density-functional theory\footnote
{Recently it has been shown that the density-functional
theory employed by M$^c$ Carley and Ashcroft (1994)
is unable to predict the existence
of certain high density, mechanically stable,
HS solids (Tejero 1997,1998).}.
Nonetheless, the crystal
density range shown in fig. 5 is enough to perform our comparison
of relative stabilities. The Gaussian widths of both solid phases are
shown in table 3, again we observe that the small spheres in the
quasicrystal are less localized than the large ones, but both spheres
are more
localized than the corresponding values of table 2. Regarding the crystal,
note the anomalous behavior of the last column; the loss of localization of
large spheres at increasing densities finally yields a vanishing minimum
in the $\alpha_2$ coordinate. Finally in fig. 6 we test again
the quasicrystal stability against variations of the distance $D$ between
large spheres. We observe features similar to those discussed from fig.
4 above, despite the shorter range of $D$ due to a longer large
sphere diameter.

Concluding our results, we note that the exchange of relative stabilities
between crystal and quasicrystal (cf. figs. 3 and 5) occurs at a diameter
ratio 0.83 (see table 1) when the free energies of both solids differ within
 the order of quasicrystal convergence errors, along the whole range of
packing fraction.

\tabcolsep 1cm
\begin{table}
\caption{\baselineskip 18pt
Same as  table 2 for a diameter ratio $\alpha=0.798$.
The disordered crystal is absent due to its low
packing fraction (see table 1).
\label{tab3}}
\begin{tabular}{ccccc} \hline\hline
$\eta$ &  $(\alpha_1)_q^{-\frac{1}{2}}$
& $(\alpha_2)_q^{-\frac{1}{2}}$ &
 $(\alpha_1)_c^{-\frac{1}{2}}$  &
$(\alpha_2)_c^{-\frac{1}{2}}$ \\
\hline
0.52 & 0.16 & 0.085 &  &  \\
0.53 & 0.14 & 0.078 & 0.15 & 0.14 \\
0.54 & 0.13 & 0.072 & 0.14 & 0.13 \\
0.55 & 0.12 & 0.067 & 0.13 & 0.12 \\
0.56 & 0.11 & 0.061 & 0.12 & 0.11 \\
0.57 & 0.10 & 0.057 & 0.12 & 0.12 \\
0.58 & 0.096 & 0.053 & 0.11 & 0.13 \\
0.59 & 0.088 & 0.049 & 0.11 & 0.15 \\
0.60 & 0.081 & 0.046 &  &  \\
0.61 & 0.075 & 0.043 &  & \\
0.62 & 0.069 & 0.041  & & \\
0.63 & 0.064 & 0.039  & & \\
0.64 & 0.059 & 0.037 & & \\ \hline\hline
\end{tabular}
\end{table}

\section*{\S5. SUMMARY AND CONCLUSION}
We have analyzed the stability of a HS binary quasicrystal by means of a
density-functional method. We have first focused upon a TIBM structure with
a diameter ratio 0.888 that maximizes the packing fraction.
It was found that this HS structure
turns out to be thermodynamically unstable i. e., it has a higher free
energy with respect to crystals and fluid of similar composition, being
a disordered crystal the solid phase of lowest free energy.
Next we have studied the quasicrystal free energy dependence upon the
distance between nearest large spheres, finding a minimum at
 a slightly more distance  than the original TIBM one.
In addition, we have found that a small increase of such distance allows
to improve the maximum packing fraction to a competitive value with
respect to a crystal of similar composition. In fact, it was found that
for a diameter ratio around 0.8, the quasicrystal is stable with respect
to such crystal, but it remains unstable as compared to the fluid.

In conclusion, as was pointed out by M$^c$ Carley and Ashcroft (1994),
it appears that crystalline
states are preferred for diameter ratios above 0.85 but, as was shown in the
present paper, a small decrease of diameter ratio to 0.83 leads to a favored
quasicrystalline state which, however, does not reach a coexistence with the
fluid. Such thermodynamical unstability is most probably due to the
assumed HS interparticle potential, in contrast to the Lennard-Jones  one
of the simulation performed by Roth et al. (1995).

\section*{ACKNOWLEDGMENTS}
The author is indebted to C. L. Henley and C. F. Tejero for helpful comments
and to D. M. Jezek for useful discussions.

\newpage

\begin{center}
\Large{\bf REFERENCES}
\end{center}

\noindent
ACHRAYAH, R., and BAUS, M., 1998, {\it Phys. Rev.} E, {\bf 57}, 4361.

\noindent
BOUBLIK, T., 1970, {\it J. Chem. Phys.}, {\bf 53}, 471.

\noindent
CATALDO, H. M., and TEJERO, C. F., 1995, {\it Phys. Rev.} B, {\bf 52}, 13269.

\noindent
DAANOUN, A., TEJERO, C. F., and BAUS, M., 1994, {\it Phys. Rev.} E, {\bf 50}, 2913.

\noindent
DENTON, A. R., and HAFNER, J., 1997a, {\it Europhys. Lett.,} {\bf 38}, 189;
1997b, {\it Phys. Rev.} B, {\bf 56}, 2469.

\noindent
ELSER, V., and HENLEY, C. L., 1985, {\it Phys. Rev. Lett.,} {\bf 55}, 2883.

\noindent
GOLDMAN, A. I., and KELTON, K. F., 1993, {\it Rev. Mod. Phys.,} {\bf 65}, 213.

\noindent
HENLEY, C. L., 1986, {\it Phys. Rev.} B, {\bf 34}, 797.

\noindent
HENLEY, C. L., and ELSER, V., 1986, {\it Phil. Mag.} B, {\bf 53}, L59.

\noindent
JANOT, C., 1994, {\it Quasicrystals} (Oxford: Clarendon Press).

\noindent
KRAJ\v{C}\'{I}, M., and HAFNER, J., 1992,  {\it Phys. Rev.} B, {\bf 46},
10669.

\noindent
LEBOWITZ, J. L., 1964, {\it Phys. Rev.,} {\bf 133}, A895.

\noindent
LIKOS, C. N., and HENLEY, C. L., 1993, {\it Phil. Mag.} B, {\bf 68}, 85.

\noindent
LUTSKO, J. F., and BAUS, M., 1990, {\it Phys. Rev.} A, {\bf 41}, 6647.

\noindent
MANSOORI, G. A., CARNAHAN, N. F., STARLING, K. E., and LELAND, T. W., 1971,
{\it J. Chem. Phys.}, {\bf 54}, 1523.

\noindent
M$^c$CARLEY, J. S., and ASHCROFT, N. W., 1994,
{\it Phys. Rev.} B, {\bf 49}, 15600.

\noindent
ROTH, J., SCHILLING, R., and TREBIN, H. R., 1990,
{\it Phys. Rev.} B, {\bf 41}, 2735;
1995, Ibid., {\bf 51}, 15833.

\noindent
STEINHARDT, P. J., and DI VINCENZO, D. P. (editors), 1991,
{\it Quasicrystals: The State of the Art}
(Singapore: World Scientific).

\noindent
TEJERO, C. F., 1997, {\it Phys. Rev.} E, {\bf 55}, 3720;
1998, Ibid., {\bf 58}, 5171.

\noindent
TEJERO, C. F., and CUESTA, J. A., 1993, {\it Phys. Rev.} E, {\bf 47}, 490.

\noindent
XU, H., and BAUS, M., 1992,
{\it J. Phys.: Cond. Matt.,} {\bf 4}, L663.

\newpage
\begin{figure}
        \begin{center}
\epsfxsize=7cm
        \epsfbox{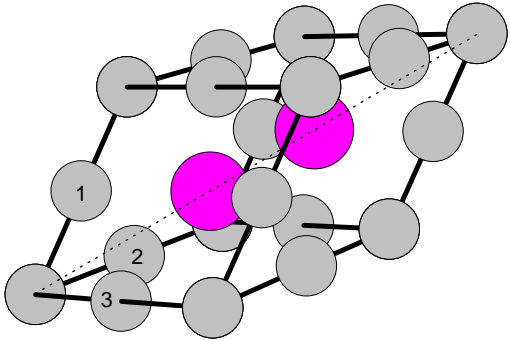}
        \caption{Decoration of a prolate rombohedron containing two large spheres located
symmetrically at both sides of the center of its long diagonal denoted by
a dotted line. The maximum diameter of small spheres can reach half a
rombohedron
edge. The separation of both large spheres is bounded by the contact with
other large spheres located in adjacent rombohedra, or eventually by the
contact with the small spheres denoted as 1, 2, 3.
}
        \end{center}
\end{figure}

\begin{figure}
        \begin{center}
\epsfxsize=7cm
        \epsfbox{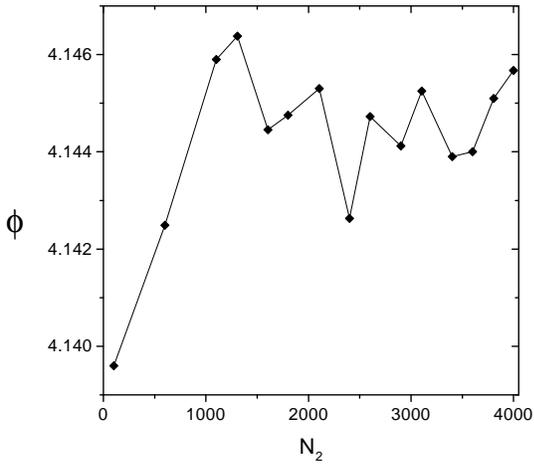}
        \caption{Free energy per unit volume $\phi=\eta[f/k_BT-3\ln(
\Lambda_2/\sigma_2)+1]$, of spherical quasicrystal ``samples'' of
different sizes
containing $N_2$ large spheres. The HS parameters are $\alpha=0.798$,
$\eta=0.55$, and the distance between both large spheres of each
prolate rombohedron is 0.627.
}
        \end{center}
\end{figure}

\begin{figure}
        \begin{center}
\epsfxsize=7cm
        \epsfbox{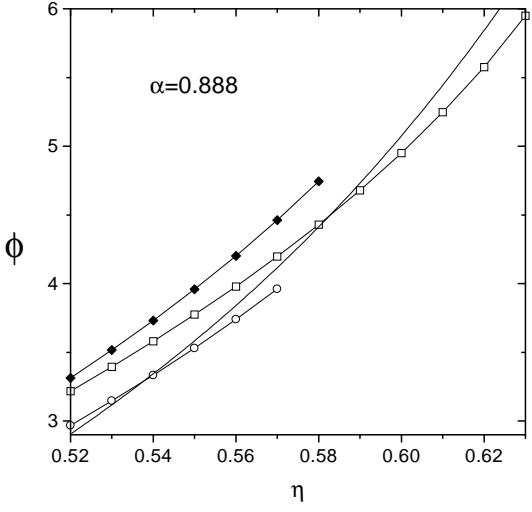}
        \caption{Free energy per unit volume  $\phi=\eta[f/k_BT-3\ln(
\Lambda_2/\sigma_2)+1]$, vs packing fraction $\eta$, for quasicrystal
(diamond points), crystal (square points), disordered crystal (circle points),
and fluid (solid line).
}
        \end{center}
\end{figure}

\begin{figure}
        \begin{center}
\epsfxsize=7cm
        \epsfbox{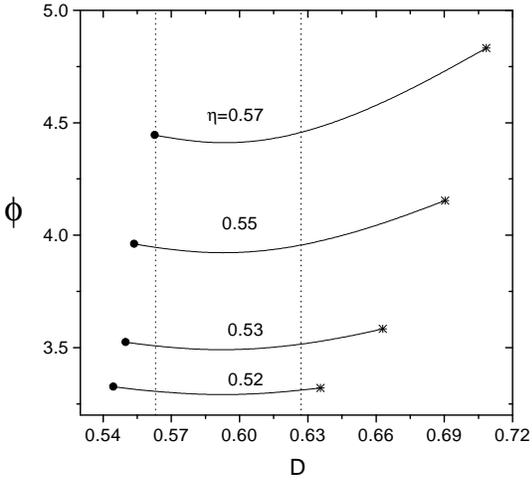}
        \caption{Quasicrystal
free energy per unit volume $\phi$, vs the distance $D$ between both large
spheres of each rombohedron, for different packing fractions. The circle
points (left end of curves) indicate that both spheres are close to
touch each other, while the star points (right end of curves) indicate
an end of mechanical stability. The dotted vertical lines correspond to
$D=0.563$ and $D=0.627$, the diameter ratio is $\alpha=0.888$.
}
        \end{center}
\end{figure}

\begin{figure}
        \begin{center}
\epsfxsize=7cm
        \epsfbox{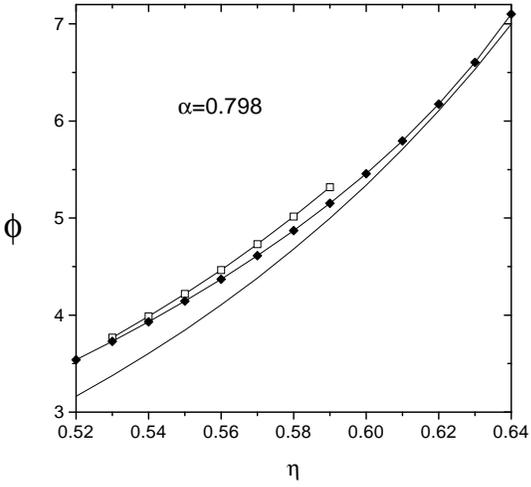}
        \caption{Same as  fig. 3. The disordered crystal is absent due to its low
packing fraction (see table 1).
}
        \end{center}
\end{figure}

\begin{figure}
        \begin{center}
\epsfxsize=7cm
        \epsfbox{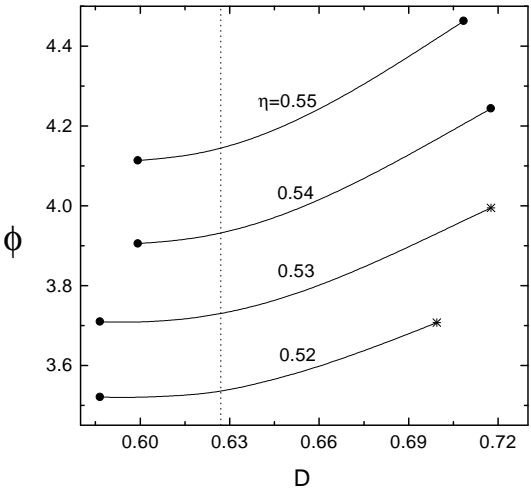}
        \caption{Same as fig. 4 for a diameter ratio $\alpha=0.798$. The circle points at
the right end of upper curves indicate that each large sphere is close to
touch the small spheres denoted as 1, 2, 3 in fig. 1.
}
        \end{center}
\end{figure}

\end{document}